\documentstyle[12pt,procsla]{article}
  
  \catcode`\@=11
  \long\def\@makefntext#1{
  \protect\noindent \hbox to 3.2pt {\hskip-.9pt  
  $^{{\ninerm\@thefnmark}}$\hfil}#1\hfill}		%CAN BE USED 
  
  \def\@makefnmark{\hbox to 0pt{$^{\@thefnmark}$\hss}}  %ORIGINAL 
  	
  \def\ps@myheadings{\let\@mkboth\@gobbletwo
  \def\@oddhead{\hbox{}
  \rightmark\hfil\ninerm\thepage}   
  \def\@oddfoot{}\def\@evenhead{\ninerm\thepage\hfil
  \leftmark\hbox{}}\def\@evenfoot{}
  \def\sectionmark##1{}\def\subsectionmark##1{}}
\newcommand{\be}{\begin{equation}}
\newcommand{\ee}{\end{equation}}
\newcommand{\bea}{\begin{eqnarray}}
\newcommand{\eea}{\end{eqnarray}}
\newcommand{\ba}{\begin{array}{l}}
\newcommand{\ea}{\end{array}}

\newcommand{\bb}{\begin{thebibliography}}
\newcommand{\eb}{\end{thebibliography}}
\newcommand{\ci}[1]{\cite{#1}}
\newcommand{\lab}[1]{\label{#1}}
\newcommand{\re}[1]{(\ref{#1})}
\newcommand{\Ds}{\displaystyle}
\newcommand{\half}{{\textstyle{\frac{1}{2}}}}

\begin{document}
  
  \centerline{\normalsize\bf AXIAL ANOMALY LOW ENERGY TESTS}
  \baselineskip=22pt
  \centerline{\normalsize\bf AND INSTANTON VACUUM MODELS}
  \baselineskip=16pt                                  
  \centerline{\footnotesize E. DI SALVO and M.M. MUSAKHANOV\footnote{Associate 
Member of ICTP; on leave from Theoretical Physics Department, 
Tashkent State University, Tashkent 700095, Uzbekistan}} 
  \baselineskip=13pt
  \centerline{\footnotesize\it Dipartimento di Fisica, Universita' di Genova}
  \baselineskip=12pt
  \centerline{\footnotesize\it Via Dodecaneso 33, 16146 Genova}
  \centerline{\footnotesize E-mail: disalvo@ge.infn.it; yousuf@ge.infn.it}
  \baselineskip=13pt
  \vspace*{0.9cm}
  \abstracts{
A low energy theorem concerning a matrix element of the QCD axial anomaly is 
tested against different instanton models. In the chiral limit the theorem is 
fulfilled by the Diakonov\&Petrov model, whereas it is violated by 
single-instanton approximations. Beyond the chiral limit the theorem, and two 
relations established for other matrix elements of the QCD axial anomaly, 
result to be violated also by the Diakonov\&Petrov model.}

\section{Introduction}

In  quantum field theory the symmetry of the classical
lagrangian may be destroyed by quantum correction. In gauge
theories the most important examples of this kind are the axial anomalies in
ElectroWeak theory and in QCD. The axial anomaly arises from noninvariance of 
the fermionic measure against axial transformations 
in the path integrals of the theory\ci{Fujikawa} (see also 
 higher-loop corrections\ci{FMP}).    
In Euclidean QCD+QED the axial anomaly reads
\be
\partial_{\mu}j^{5}_\mu 
= i N_f \frac{g^2}{16\pi^2} G\tilde G 
+
i N_c \frac{e^2}{8\pi^2} \sum_{f}{Q^{2}_{f}} F\tilde F   
+ 2\sum_{f}m_{f}\psi^{\dagger}_{f}\gamma_{5}\psi_{f},
\lab{div}
\ee
where 
$j^{5}_\mu =\sum_{f}\psi^{\dagger}_{f}\gamma_{\mu}\gamma_{5}\psi_{f}\, 
(f=u,d,s)$ 
is the quark singlet axial current, $\psi_{f}$ the quark field,
 $N_f$ the number of the flavors, $g$ the QCD coupling constant,  
 $2G\tilde G= \epsilon^{\mu\nu\lambda\sigma}
 G^{a}_{\mu\nu}  G^{a}_{\lambda\sigma}$, $G^{a}_{\mu\nu}$ 
the gluon field strength operator, 
$N_c$ the number of colours, $e$ the QED coupling constant, $Q_{f}$ the 
electric fractional charges of the quarks, $F_{\mu\nu}$ 
the photon field strength operator.
We have explicitly included
the contributions of the current masses of light quarks $m_{f}$. 
It was pointed out\ci{Shi} that this equation leads to a  
nontrivial low-energy theorem ($LET$) and to useful relations. 

In the present paper we  consider the matrix element of eq. \re{div} $(i)$ 
between vacuum and two-photons states and $(ii)$ between vacuum and meson 
states.

Analysis\ci{RR96} of case $(i)$, and nonvanishing of the $\eta^{'}$ meson mass
even in chiral limit due to axial anomaly, leads to the conclusion that in the 
kinematical configuration of zero virtualities of each line  in the
form factor the matrix element of the divergence of the singlet axial current 
vanishes, implying the following $LET$:   
\be
 \langle 0|N_f \frac{g^2}{16\pi^2}G\tilde G | 2\gamma \rangle =
 N_c \frac{e^2}{4\pi^2} \sum_{f} Q^{2}_{f} 
F^{(2)}\tilde F^{(3)} 
+ 2i\sum_{f} m_{f}
\langle 0|\psi^{\dagger}_{f}\gamma_{5}\psi_{f}| 2\gamma \rangle,
\lab{theorem}  
\ee
where $F_{\mu\nu}^{(i)}= \epsilon_{i,\mu}q_{i,\nu} - \epsilon_{i,\nu}q_{i,\mu}$
and $q_i ,\, \epsilon_{i}\,(i=2,3)$ are respectively the momenta and
polarization vectors of photons.

Eq. \re{theorem}   gives the exact answer for the matrix element of the gluonic 
operator.
Of course gluons can couple to photons only by virtual quarks.
In perturbation theory the left side of the eq. \re{theorem} is
at least $O(g^{4}e^{2})$, whereas the right side of this equation does not
contain any strong coupling at all. It is evident that the $LET$ can be
fulfilled only out of the framework of perturbation
theory. The factor $g^2$ at the left side must be completely cancelled by some
nonperturbative contribution, such as the one, of order $g^{-2}$, provided by 
instantons\ci{BPST75}.
The computation of the left side of eq. \re{theorem} amounts to calculating the 
instanton contribution in the Euclidean space to the three-point function
\be
\tau _{\mu \nu}(x_1 , x_2 , x_3 ) =
\langle 0| T[{g^2}G\tilde G(x_1) j^{em}_\mu (x_2 )
j^{em}_\nu (x_3 )] |0 \rangle , 
\label{corr}
\ee  

($j^{em}_{\mu}$ being the electromagnetic current)
which contains  needed  information on the matrix element
$\langle 0|G\tilde G | 2\gamma \rangle$ and also  
useful physical information concerning $\eta \rightarrow 2 \gamma$  decay, 
$2 \gamma$- transitions in heavy quarkonia etc.

The matrix elements of eq. \re{div} between vacuum and meson states
lead to the relations 
\begin{eqnarray}
 \langle 0|N_f \frac{g^2}{16\pi^2}G\tilde G | \eta \rangle &=&
 2 i m_{s}\langle 0|\psi^{\dagger}_{s}\gamma_{5}\psi_{s} | \eta \rangle, \ 
\lab{relation1}    
\\
\langle 0|N_f \frac{g^2}{16\pi^2}G\tilde G | \pi^{0} \rangle &=&
i (m_{u}-m_{d})
\langle 0|\psi^{\dagger}\tau_{3}\gamma_{5}\psi | \pi^{0} \rangle,
\lab{relation2}
\end{eqnarray}

which provide, together with $LET$, more stringent tests for instanton models.

Instantons - whose presence in QCD is a well established fact, at least at a 
phenomenological level and in numerical simulations of QCD vacuum\ci{SS96} -
constitute the main input to our calculations, especially in connection with 
quark propagation. 
To our present knowledge the instanton structure of QCD vacuum is 
concentrated in an average size $\rho$ and in an average interinstanton 
distance $R$, such that\ci{DP84,Shu82}
\be
\rho =1/3 fm, \, \, \, \,\,\,\, R=1fm .
\label{rho,R}    
\ee
Therefore the packing parameter $(\rho /R)^4 = 0.012$ is small, legitimating
independent averaging over positions and orientations of the instantons;
moreover in some cases it is even possible to  use  
the quark propagator in the single-instanton field\ci{BCCL78}. 

In a previous work\ci{MK} $LET$ was used as a test for the Diakonov-Petrov 
($DP$) ansatz\ci{DP86,DPW95} of the low-energy 
QCD effective action in the chiral limit. The anzatz, based on an 
interpolation formula for the quark propagator in the field of a single 
instanton, does satisfy $LET$ to $\sim 17\%$ accuracy. 
             
One of the aims of our present work is to formulate more stringent tests of 
the $DP$ model, by 
taking explicitly into account the contribution of the current quark masses 
$m_{f}$. We also test the validity of different instanton 
vacuum models by calculating $\tau_{\mu\nu}$ 
in the range of small virtualities and testing it against $LET$.

In  section 2  we calculate $\tau_{\mu\nu}$ in single-instanton 
approximation($SIA$). 
In  section 3 we rederive the $DP$ effective
action starting from results of Lee$\&$Bardeen\ci{LB79} ($LB$)
and accounting for current quark masses $m_{f}$.
In section 4 we calculate $\tau_{\mu\nu}$ by this effective action 
and check it by $LET$. In section 5 we calculate in the same framework both 
sides of relations \re{relation1} and \re{relation2} and compare each left side
with its respective right side.
 
\section{Single-instanton approximation}

A lot of papers have been devoted to instanton calculus to
single-instanton approximation (see for example the calculations of the
correlator of two vector currents in\ci{CL78,AG78,BEGZ78}). 
In this approximation  the calculation of $\tau_{\mu\nu}$ 
amounts to computing integrals like
\be
\ba \Ds
\int d^{4}z_{\pm}dU_{\pm}\langle  T(g^{2}G\tilde G(x_{1}) j^{em}_{\mu} (x_{2} )
j^{em}_{\nu} (x_{3} ) \rangle_{\pm} = 
\\ \Ds
 \sum_{f}{Q^{2}_{f}}\int d^{4}z_{\pm} (G\tilde G(x_{1})_{\pm}
Tr(\gamma_{\mu}S_{\pm}(x_{2},x_{3})\gamma_{\nu}S_{\pm}(x_{3},x_{2})),
\label{1-inst}
\ea
\ee

where $z_{\pm}$ and $U_{\pm}$ are the position and orientation of the 
(anti-)instanton, assuming instanton integration sizes to be peaked at  
$\rho$, eq. \re{rho,R}.
Here 
\be
(g^{2}G\tilde G(x))_{\pm} = \pm f(x-z)=
\pm\frac{192 \rho^4}{\left[ \rho^2 + (x - z)^2 \right]^4}
\lab{GtildeG}   
\ee
and $S_{\pm}=(i\hat D_{\pm} + im)^{-1}$ \ ~ \ is the full quark propagator \ ~ \
in presence of a single (anti)instanton. 

Starting from the expression of $S_{\pm}$ given in ref.\ci{BCCL78}, we obtain

\be
\ba \Ds
\tau _{\mu \nu}(x_1 , x_2 , x_3 ) =   
\frac{N}{V} \frac{192N_{c}\rho^{6}}
{3 \pi^{4}}\sum_{f}{Q^{2}_{f}}   
\int d^{4}z  h^{4}(x_{1} - z)
\frac{h(x_{2}-z)h(x_{3}-z)}{(x_{2}-x_{3})^{2}}
\\  \Ds
\times
[\frac{h(x_{2}-z) + h(x_{3}-z)}{(x_{2}-x_{3})^{2}}
+ h(x_{2}-z)h(x_{3}-z)]
\epsilon_{\mu\nu\alpha\beta} \ ~ (x_{2}-z)_{\alpha}(x_{3}-z)_{\beta},
\label{single-i-corr}
\ea
\ee   
where
$h(x) = (\rho^{2} + x^{2})^{-1} .$ 
Then the correlator, that is the Fourier transform $\hat{\tau}_{\mu\nu}$ of the
three-point function $\tau_{\mu\nu}$, to be tested against $LET$, results to be

\be\ba\Ds
\hat{\tau}_{\mu\nu}(q_{1},q_{2},q_{3}) = 
\frac{N}{V} \frac{N_{c}\rho^{2}}
{3 \pi^{4}}\sum_{f}{Q^{2}_{f}}   
(2\pi)^{4}\delta(\sum_{i}{q_{i}}){\hat f}(q^{2}_{1})
\\ \\ \Ds
\times
4!\int_{0}^{1}ada\int_{0}^{1}db(a(1-a+ab(1-b))^{-3/2}
\epsilon_{\mu\nu\alpha\beta}
(-\frac{\partial^{2}}{\partial p_{2,\alpha}\partial p_{3,\beta}}) J(P),
\label{8-int} 
\ea\ee
as can be shown by applying the Feynman integration technique. We have set

\be\ba\Ds
{\hat f} (q^2_1) = \int d^4 x_1\, \frac{192\rho^4}{(\rho^2+x_1^2)^{4}}, 
\, \exp(iq_1x_1), 
\\ \\ \Ds
J(P)=\int d^{8}Y\, [Y^{2}+ r^{2}]^{-5} \, \exp{iPY},
\lab{J}
\ea\ee
having introduced the 8-dim vectors $P=(p_{2},p_{3})$, 
$Y=(y_{2},y_{3})$, with
$$p_{2}=\frac{q_{2}+bq_{3}}{(1-a + ab(1-b))^{1/2}}, \, \, 
p_{3}= \frac{q_{3}}{a^{1/2}}, \,\, r^{2}=\rho^{2}(1-ab).$$
The latter integral \re{J} can be easily calculated 
and is reduced to the MacDonald function,  i. e.,
\be
J(P)= \frac{\pi^{4}}{4!}\frac{P}{r}K_{1}(Pr) ,  \,\, P=(P^{2})^{1/2} .
\lab{J(P)}
\ee

In the limit of small momenta
\be
K_{1}(Pr)=\frac{1}{Pr} + \frac{Pr}{2}[ \ln {\frac{Pr}{2}} + C - 1/2 ]
+...,
\lab{K1}
\ee
where $C = 0.577215...$ is the Euler constant.

Eqs. \re{8-int}, \re{J(P)}, \re{K1}  imply that
$\hat{\tau}_{\mu\nu}(q_{1},q_{2},q_{3})$ is divergent for 
$q_{i}^2 \rightarrow 0$, that is, the model badly violates $LET$. Nor does 
agree with $LET$ the
improved single-instanton approximation\ci{SS96}, where the zero-mode 
contribution to the propagator has been modified by replacing $m \rightarrow 
m^{*}$, where $m^{*}$ is the effective mass of the quark accounting for the 
influence of the surrounding instantons).
We conclude that single-instanton approximation badly violates $LET$, therefore
it is not suitable for  
calculating three-point functions at small momenta. By the way we note that
two-point functions of vector currents are not so sensitive to large distance 
effects. 
The violation of $LET$ - and in particular the divergence in massless quark
aproximation - is related to the slow decrease of $\tau_{\mu\nu}$ at large 
distances. Indeed we have neglected rescattering effects of quarks by other 
instantons, which, during quark propagation, lead to the formation of the 
constituent quark, producing a suitable effective mass and providing needed 
exponential decrease with distance. 
Such effects can be described by an effective action.
 
\section{A low energy QCD effective action}

It is natural to choose the singular gauge for the instantons in describing
many instanton effects in the propagation of the quarks. In the case of
a small packing parameter  it is possible to do the following ansatz for the 
background instanton field:
\be
A_{ \mu}(x)= \sum_{+}^{N_+} A_{+ , \mu}(x; \xi_{+}) + 
\sum_{-}^{N_-} A_{-, \mu}(x; \xi_{-}), \, 
(\xi_{\pm}=(z_{\pm}, U_{\pm},\rho_{\pm})),
\ee
where, $z_i$, $U_i$ and $\rho_i$ the position, orientation and size of the
$i$-th instanton.
The canonical partition function of the
 $N_+$instantons and $N_-$ antiinstantons can be schematically
written as
\be
Z_{N_+ , N_-} = \int det_N \exp( - V_g )\prod_{i}^{N_+ , N_-} , 
d^4 z_i dU_i dn(\rho_i),
\ee                                                      
where $V_g$ is the instanton-(anti)instanton interaction potential generated
by the gluon field action and $det_N$ is a quark determinant in the instanton 
field. The main assumption of the instanton model is  that
$V_g$ is repulsive  at small distances between instanton and antiinstanton.
This should 
provide the stabilization of the instanton sizes and of the interinstanton
distances, as  discussed in the introduction. 
We mainly deal with $det_N ,$ 
which describes the influence of light quarks.

Lee\&Bardeen\ci{LB79} ($LB$) calculated the quark propagator in a more 
sofisticated approximation than $SIA.$ In particular they found that
\be
det_{N}=det B, \,\,
B_{ij}= im\delta_{ij} + a_{ji}, 
\lab{B}
\ee
where $a_{ij}$ is the overlapping matrix element of the quark zero-modes 
$\Phi_{\pm , 0} $ generated by instantons.
This matrix element is nonzero only between instantons and antiinstantons
(and vice versa) due to the chiral factor in $\Phi_{\pm , 0} $ , i. e.,
\be
a_{-+}=-<\Phi_{- , 0} | i\hat\partial |\Phi_{+ , 0} > .
\lab{a}
\ee
The overlapping of the quark zero-modes causes quark jumping from one instanton 
to another one during propagation.

It is clear from \re{B} that 
for $N_+ \not= N_-$ $det_N \sim m^{|N_+ - N_-|}$, so the fluctuations of  
$|N_{+}-N_{-}|$  are strongly suppressed due to presence of light quarks. 
Therefore we assume $N_{+}=N_{-}=N/2 .$

Following the procedure suggested in ref.\ci{Tokarev}, we get 
the fermionization of $LB$'s result, 
i.e.,
\be
\ba  \Ds
det_N = \int D\psi D\psi^{\dagger} \exp(\int d^4 x
\sum_{f}\psi_{f}^{\dagger}i\hat\partial \psi_{f})     \\   \Ds
\times \prod_{f}(\prod_{+}^{N_{+}}(im_{f} + V_{+}[\psi_{f}^{\dagger} ,\psi_{f}])
\prod_{-}^{N_{-}}(im_{f} + V_{-}[\psi_{f}^{\dagger} ,\psi_{f}])) ,
\label{part-func}
\ea
\ee
where 
\be
V_{\pm}[\psi_{f}^{\dagger} ,\psi_{f}]= 
\int d^4 x (\psi_{f}^{\dagger} (x) i\hat\partial
\Phi_{\pm , 0} (x; \xi_{\pm}))
\int d^4 y 
(\Phi_{\pm , 0} ^\dagger (y; \xi_{\pm} )  
i\hat\partial \psi_{f} (y)).
\ee
Eq. \re{part-func} coincides with the ansatz 
for the fixed $N$ partition function
postulated by $DP$, except for the sign in front of 
$V_{\pm}$.  Keeping in mind the low density of the instanton media, which 
allows independent averaging over positions and orientations 
of the instantons, eq. \re{part-func} leads to the partition function

\be
 Z_N = \int D\psi D\psi^\dagger \exp  (\int d^4 x \, \psi^\dagger 
i \hat\partial  \psi )  \,  W_{+}^{N_+}  \, W_{-}^{N_-}, 
\lab{Z_NW}
\ee
where
\be\ba\Ds
W_\pm =\int d^4 \xi_{\pm}\prod_{f} (V_{\pm}[\psi_{f}^{\dagger} 
\,\psi_{f}] + i m_{f}) =
\\ \Ds
(-i)^{N_{f}}\left(  \frac{4\pi^2\bar\rho^2}{N_c} \right)^{N_f}
\int \frac{d^4 z}{V} 
\det_{f}(i J_\pm (z) - \frac{m ~ N_c}{4\pi^2 \rho^2})
\ea\ee

and

\be
J_\pm (z)_{fg} = \int \frac {d^4 kd^4 l}{(2\pi )^8 } \exp ( -i(k - l)z)
\, F(k^2) F(l^2) \, \psi^\dagger_f (k) \half (1 \pm \gamma_5 ) \psi_g (l) .
\lab{J_pm}
\ee
The form factor $F$ is related  to the zero--mode wave function 
 in momentum space $\Phi_\pm (k; \xi_{\pm}) $ and is equal to
\be
F(k^2) = - t \frac{d}{dt} \left[ I_0 (t) K_0 (t) - I_1 (t) K_1 (t)
\right], \,\, t =\frac{1}{2} \sqrt{k^2} \bar\rho.
\ee
 
\section{Calculation of the correlator by the $DP$ effective action}

In quasiclassical (saddle point) approximation  any gluon operator derives its 
main contribution from instanton background. 
In the following   the operator $g^2 G\tilde G(x)$  will be considered.
Owing to the low density of the instanton medium, it is possible to neglect  
the overlap of the fields of different instantons. In that case,  
the matrix element of $g^2G\tilde G(x)$ with any other 
quark operator $Q$ is
\be \ba
\langle {g^2}G\tilde G(x) Q \rangle_N = 
Z_{N}^{-1} \int D\psi D\psi^\dagger \exp  (\int d^4 x \psi^\dagger i 
\hat\partial \psi)   
\\ \\
 \times
\left( N_{+} \left( W_{G\tilde G +} (x) Q \right) W_{+}^{N_+ - 1}   
W_{-}^{N_-}  +  N_{-} \left( W_{G\tilde G -} (x) Q \right) W_{+}^{N_+ }   
W_{-}^{N_- - 1} \right) ,
\lab{GtildeGQ1}     
\ea \ee
where
\be
W_{G\tilde G \pm} = \pm\left(  \frac{4\pi^2\bar\rho^2}{N_c} \right)^{N_f}
\int \frac{d^4 z}{V}\, f(x-z) \, 
\det_{f}( J_\pm (z) + i\frac{m ~ N_c}{4\pi^2 \rho^2}) .
\lab{W_GtildeGQ}
\ee 
 It is useful to introduce the external fields $\kappa (x)$, coupled to
$g^2 G\tilde G$, and $a$, such that $\hat D = \hat\partial  - i e Q_{f} 
\hat a .$ Starting from \re{W_GtildeGQ}, we find  that the partition 
function $\hat Z[\kappa , a]$  
describing mesons\ci{MK} in presence of such external fields is
\be
\hat Z[\kappa , a]=  \int D\Phi_{+}D\Phi_{-} exp (-W[\Phi_+,\Phi_-]),  
\label{intz}
\ee
where
\be\ba\Ds
W[\Phi_+,\Phi_-] = \int d^4 x (w_a + w_b -w_c),
\\ \\ \Ds
w_a = (N_{f}-1)\frac{N}{2V}(\prod_{f}M_{f}^{-1}det \Phi_{+})^{(N_{f}-1)^{-1}}
\ + (+ \rightarrow -), 
\\ \\ \Ds
\ w_b = \frac{N_c}{4\pi^{2}\rho^{2}}Tr(m ~ (\Phi_{+}+\Phi_{-})),
\\ \\ \Ds
w_c = \sum_{f} Tr ln \frac{i \hat \partial +i F^2 ~ (\Phi_+ A_+ + \Phi_- A_-)}
{i \hat \partial + i m_f},
\\ \\ \Ds
A_{\pm} = \left[ \left( 1 \pm ( \kappa  f) \right)^{N_{f}^{-1}} \right]
\frac {1}{2}(1 \pm \gamma_5).
\\
\lab{Zkappa2}    
\ea
\ee
The saddle point  of the integral \re{intz} is located at 
$(\Phi_{\pm})_{fg} = M_{f}\delta_{fg}$, a self-consistency equation for the 
effective quark mass, i. e.,
\be
4 N_c V \int \frac{d^4 k}{(2\pi )^4} 
\frac{M_{f}^{2} F^4 (k^2)}{M_{f}^{2} F^4 (k^2) + k^2}
=  N  + \frac{m_{f}M_{f}VN_{c}}{2\pi^{2}\rho^{2}},
\lab{selfconsist}  
\ee
being imposed. Eq. \re{selfconsist} describes also the shift of the effective
mass of the quark $M_f$ due to current mass $m_f$.
Expanding \re{selfconsist} with respect to $m_f$, we have
 $M_{f}=M_{0}+\gamma  m_{f},$
where 
\be
\gamma^{-1} \, = \,
\rho^2\int_{0}^{\infty}d k^{2} 
\frac{k^4 F^4 (k^2)}{(M_0^2 F^4 (k^2) + k^2 )^2}.
\lab{gamma}
\ee
Such integrals converge due to the form factor $F(k^2).$ 
Assuming for the instanton model parameters $\rho$ and $R$ the values 
\re{rho,R} - which correspond to $M_{0} \,=\, 340\, MeV$ - we find
\be
 \gamma\, = \, 2.4 ~ .
\lab{gamma1}
\ee
The quark condensate  is then given by 
\be
- V^{-1}Z_{N}^{-1}\frac{dZ_{N}}{dm}|_{\kappa = 0} 
 \,=\,- \frac{N_{c}M_{0}}{2\pi^{2}\rho^{2}}
\,=\,- (265 \, MeV)^{3} \, \sim N_{c}\frac{1}{\rho R^{2}}.
\lab{condensate1}
\ee
This quantity can be also calculated by formula\ci{DP86} 
$-i <\psi^{\dagger}\psi >_{Euclid}\,= \, iTr\,S$, yielding
\be\ba\Ds
-i <\psi^{\dagger}\psi >_{Euclid} \,=\, - 4N_{c}\int \frac{d^4 k}{(2\pi )^4} 
\frac{M_{0} F^2 (k^2)}{M_{0}^{2} F^4 (k^2) + k^2} 
\\ \Ds
\,=\,- (255 \, MeV)^{3}   \, \sim N_{c}\frac{1}{\rho R^{2}}. 
 \label{condensate2}
\ea\ee
Although coming from different formulas, predictions \re{condensate1} and 
\re{condensate2} have the same parametric dependence and agree with the 
phenomenological value, i. e., 

$$-i <\psi^{\dagger}\psi >_{Euclid} = -(240-250\, MeV)^3.$$
The three-point function can be derived from the functional relation 
\be
\tau_{\mu\nu} (x_1 , x_2 , x_3 ) = \frac {\delta \hat W  [\kappa , a] }
{\delta \kappa (x_1 ) \delta a_\mu (x_2) \delta a_\nu (x_3) } 
|_{ \kappa , a = 0},
\lab{tau}
\ee 
where\ci{MK}
\be 
\hat W  [\kappa , a] = 
 \sum_{f} Tr ln 
\frac{ i\hat D + iM_{f}F^2 ~ (A_+ + A_-)}{i\hat\partial + i m_f}.
 \lab{hatW}
 \ee 

It is clear from eq. \re{hatW} that the vertex factors in the diagram of the
process are $eQ_{f}\gamma_{\mu}$ and $i M_{f} f F^{2}N_{f}^{-1}\gamma_{5}$.
Taking the Fourier transform of \re{tau}, we get
\be
\hat{\tau}_{\mu\nu}(q_{1},q_{2},q_{3}) = \hat f(q^{2}_{1} )
N_c  e^2 \sum_{f}{Q^{2}_{f}} 
8M_{f}^2 \epsilon^{\mu\nu\lambda\sigma}q_{2\lambda}q_{3\sigma}
\Gamma_{f} (q^{2}_{1},q^{2}_{2},q^{2}_{3} ),
\lab{tau1}  
\ee
where $\Gamma_{f} (q^{2}_{1},q^{2}_{2},q^{2}_{3} )$, the form factor coming 
from the diagram of the process considered, may be calculated analytically if 
we approximate the form factor $F$ by $1.$ 
As a result,  the left side of eq. \re{theorem} reduces to
\be
(N_f \frac{g^2}{16\pi^2})(\frac {4e^2 N_c}{g^2 N_f} \sum_{f}{Q^{2}_{f}} )
F^{(2)}\tilde F^{(3)}    ,
\lab{theorem2}
\ee
which coincides with the first term at the right side of  eq. \re{theorem}. 
If we take into account the form factor $F$ in \re{tau1}, and give the model
parameters the values \re{rho,R}, in the chiral limit we find\ci{MK}
a variation of $\sim 17\%$. Beyond the chiral limit the left side of 
\re{theorem} receives the contribution 
\be
- \frac{N_c  e^2}{2\pi^{2}} \sum_{f}{Q^{2}_{f}} 
0.2\gamma \rho m_{f} 
\epsilon^{\mu\nu\lambda\sigma}q_{2\lambda}q_{3\sigma}.
\lab{Om_f} 
\ee
This quantity has to be compared with the explicit contribution of the current 
quark masses to the right side of \re{theorem},
which, too, may be calculated by formulas \re{hatW} and \re{tau} substituting
$iM_{f} f F^{2}N_{f}^{-1}$ by $2im_{f}$ in the $\gamma_5$ vertex.
Approximating again $F \sim 1$, we obtain
\be
2i\sum_{f} m_{f}\langle 0|\psi^{\dagger}_{f}\gamma_{5}\psi_{f}| 
2\gamma \rangle = \frac{N_c  e^2}{2\pi^{2}} \sum_{f}{Q^{2}_{f}} 
\frac{m_{f}}{M_{f}} \epsilon^{\mu\nu\lambda\sigma}q_{2\lambda}q_{3\sigma}. 
\lab{mem_f1}      
\ee
The ratio of \re{Om_f} to \re{mem_f1} at the parameter values \re{rho,R}  
results to be
\be
-\,0.2\,\gamma\,\rho\, M_{0}\,=\, - \, 0.28 
\ee
and not 1, as demanded by $LET$. We stress the neat contradiction
with the theorem, not only in magnitude but also in the sign. 
So the $DP$ model (eq. \re{Z_NW}) 
fails to reproduce $LET$ beyond chiral limit.

\section{Other refined tests for instanton models}

The matrix elements of the anomaly \re{div} between vacuum and $\eta$-meson
or $\pi^{0}$
states lead to more stringent tests of the $DP$ model.

The partition function   \re{Zkappa2}
 describes mesons with matrices $\Phi_{\pm}$, whose usual decomposition
is\ci{ANVZ89}          
\be
\Phi_{\pm} = \exp(\pm \frac{i}{2}\phi )M\sigma \exp(\pm \frac{i}{2}\phi ) ,
\lab{mesons}
\ee
$\phi$ and $\sigma$ being $N_{f} \times N_{f}$ matrices.
The saddle-point condition demands $\sigma = 1,\, \Phi_{\pm} = 0$. The usual 
decomposition for the  pseudoscalar fields 
$\phi = \sum_{0}^{8}\lambda_{i}\phi_{i}$   
may be used, where $Tr \lambda_{i}\lambda_{j} = 2\delta_{ij}$ and
$\phi_{8(3)}$ can be identified with the $\eta(\pi^{0})$-meson state.

Firstly we consider the matrix element in which the $\eta$-meson is involved.
Neglecting the small admixture factor ($\sim 0.1$) with the pure singlet 
state\ci{Ball95}), the matrix element of the divergence \re{theorem} 
between the $\eta$-meson and the vacuum leads to eq. \re{relation1}, which 
can be used as a test for instanton models.

As it is clear from previous considerations, the factor $g^{2}G\tilde G$
generates the vertex $i M f F^{2}\gamma_{5}N_{f}^{-1}$ and the $\eta$-meson
gives rise to $i M \lambda_{8} F^{2}\gamma_{5}.$ The structure of the mass 
matrix is 
$$
M \, = \, M_{0} \, + \, \gamma (m_{s}(\frac{1}{3} -\frac{1}{\surd 3}\lambda_{8})
+ m_{u}\frac{1+\tau_{3}}{2} + m_{d}\frac{1-\tau_{3}}{2}).
$$ 
Then at small $q$
\be
 \langle 0|N_f \frac{g^2}{16\pi^2}G\tilde G | \eta \rangle   = 
2\gamma m_{s}(-\frac{1}{\surd 3}tr(\lambda_{8})^{2}) 4 N_{c}
\int \frac{d^4 k}{(2\pi )^4} 
\frac{M_{0} F^4 (k^2)}{M_{0}^{2} F^4 (k^2) + k^2}.
\lab{me1}
\ee                                               
Applying the same procedure that led to eq. \re{selfconsist}, we get 
\be
 \langle 0|N_f \frac{g^2}{16\pi^2}G\tilde G | \eta \rangle   = 
2\gamma m_{s}(-\frac{2}{\surd 3})\frac{N}{V M_{0}} \,
\sim m_{s}\frac{N_c^{1/2}}{\rho R^{2}} .                  
\lab{me11}  
\ee
The right side of \re{relation1} is 
\be
2m_{s}\langle 0|\psi^{\dagger}_{s}\gamma_{5}\psi_{s} | \eta \rangle  =
2 m_{s} (-\frac{2}{\surd 3})4 N_{c}
\int \frac{d^4 k}{(2\pi )^4} 
\frac{M_{0} F^2 (k^2)}{M_{0}^{2} F^4 (k^2) + k^2}\,
\sim m_{s}\frac{N_c^{1/2}}{\rho R^{2}} .                  
\lab{me2}
\ee                                                
On the other hand eq. \re{condensate2} yields
\be
2m_{s}\langle 0|\psi^{\dagger}_{s}\gamma_{5}\psi_{s} | \eta \rangle  =
2 m_{s} (-\frac{2}{\surd 3}) i <\psi^{\dagger}\psi > .
\lab{me21}
\ee
Now let us confront such equations with some consequences of the relation
\re{relation2}, where the $\pi^0$-meson is involved.
Repeating for this case the calculations applied to relation \re{relation1}, 
the left side of equation \re{relation2} yields
\be
\langle 0|N_f \frac{g^2}{16\pi^2}G\tilde G | \pi^{0} \rangle   = 
2\gamma (m_{u}-m_{d})\frac{N}{V M_{0}} ,   
\lab{me3}  
\ee
while the right side results in
\be
2i(m_{u}-m_{d})\langle 0|\psi^{\dagger}\frac{\tau_{3}}{2}
\gamma_{5}\psi | \pi^{0} \rangle  =
2 (m_{u}-m_{d}) i <\psi^{\dagger}\psi >. 
\lab{me4}   
\ee
The ratio of \re{me11} to \re{me21} equals the ratio of \re{me3} to \re{me4} 
and at the parameter values \re{rho,R} yields
\be
\frac{N}{V M_{0}i <\psi^{\dagger}\psi >}\, =\, 0.66
\ee
and not 1, as demanded  by relations \re{relation1} and
\re{relation2}. Again the $DP$ model (eq. \re{Z_NW}) strongly contradicts some
consequences of the operator equation \re{div} beyond the chiral limit.
   
\section{Conclusions}

We recall the main results of our treatment.
$LET$ and relations \re{relation1}-\re{relation2}  constitute sensitive tests 
for instanton models.  The single-instanton 
approximation\ci{CL78},\ci{AG78},\ci{BEGZ78} 
and its modifications\ci{Shu82},\ci{SS96} badly violate $LET$.
On the contrary the $DP$ model, which takes into account the multiinstanton
effects in the propagation of the quarks, agrees satisfactory with this theorem
in the chiral limit. However, beyond such a limit, even this model violates
$LET$ and the relations \re{relation1}-\re{relation2}. 

~~~~~~~~~~~~~~~~~~~~~
              
\centerline {Acknowledgements}

~~~~~~~~~~~~~~~~~~~~

The  work of M. Musakhanov is supported in part  by the NATO-CNR grant 
and by grant INTAS-93-239ext. One of the authors (M.M.) is very grateful to 
theDepartment of Physics of the University of Genova for its hospitality.

\end{document}